\documentclass[11pt,a4paper]{article}

\usepackage[T1]{fontenc}
\usepackage{lmodern}
\usepackage{amsmath,amssymb,amsthm}
\usepackage{graphicx}
\usepackage{algorithm}
\usepackage{algpseudocode}
\usepackage{hyperref}
\usepackage{xcolor}

\newtheorem{theorem}{Theorem}
\newtheorem{lemma}[theorem]{Lemma}

\theoremstyle{definition}
\newtheorem{definition}[theorem]{Definition}

\theoremstyle{remark}

\title{Polyominoes Folding into Four and Five Distinct Boxes}

\author{Erik D. Demaine \thanks{Massachusetts Institute of Technology, USA \texttt{edemaine@mit.edu}}
    \and
    Toshihiro Shirakawa \thanks{\texttt{70ku51n@gmail.com}}
    \and
    Conner Silverstein \thanks{\texttt{connersilverstein@gmail.com}}
    \and
    Reed Silverstein \thanks{\texttt{reedsilverstein@gmail.com}}
    \and
    Michael Tardibuono \thanks{Alum of the University of Waterloo, Canada \texttt{michael.tardibuono.math@gmail.com}}
    \and
    Ryuhei Uehara \thanks{Japan Advanced Institute of Science and Technology, Japan \texttt{uehara@jaist.ac.jp}}
}

\date{}

\begin{document}
\maketitle

\begin{abstract}
We present new results on polyomino nets that fold into four or five distinct boxes.
With the help of AI, we first found examples with areas 532 and 2128, respectively.
We analyze these examples and reduce their areas to 178 and 358, respectively.
We also generalize the construction of the five-box net and show that 
infinitely many polyomino nets fold into five distinct boxes.
\end{abstract}

\section{Introduction}

\begin{figure}[tbhp]
\centering
\begin{minipage}{0.58\textwidth}
 \centering
 \includegraphics[bb=0 0 322 184,width=0.85\linewidth]{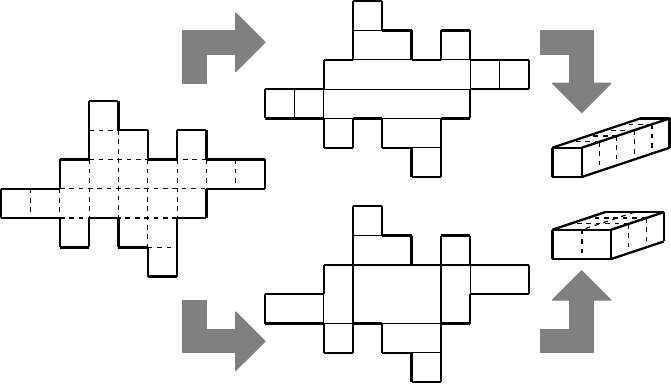}
 \caption{A polygon that folds into two boxes of size $1\times 1\times 5$ and $1\times 2\times 3$ in \protect\cite{MitaniUehara2008}.}
 \label{fig:sample}
\end{minipage}
\end{figure}

In computational origami, the relationship between a polygon $P$ and a polyhedron $Q$ folded from $P$ is one of the most important topics.
A natural restriction in computational studies is to consider orthogonal polygons and polyhedra made up of unit squares and unit cubes, respectively. These are known as \emph{polyominoes} and \emph{polycubes}, respectively.
(In this paper, a box is a rectangular cuboid with integer side lengths unless stated otherwise.)
One interesting problem in this area asks whether a polyomino can fold into multiple pairwise noncongruent boxes (see \figurename~\ref{fig:sample} for a simple example).
Biedl et al.~\cite{BiedlChanDemaineDemaineLubiwMunroShallit1999} first found a polyomino of area 22 that folds into two distinct boxes of sizes $1\times 1\times 5$ and $1\times 2\times 3$. They also found a polyomino of area 34 that folds into two distinct boxes of sizes $1\times 1\times 8$ and $1\times 2\times 5$.
(See the books on geometric folding algorithms by Demaine and O'Rourke~\cite[Chapter 25.9]{DemaineORourke2007}.)
Mitani and Uehara proved that infinitely many polyominoes fold into two distinct boxes \cite{MitaniUehara2008}.
Later, Shirakawa and Uehara showed that infinitely many polyominoes fold into three distinct boxes \cite{ShirakawaUehara2013}.
The smallest example in their work had area $532$ and folded into three boxes of sizes $7\times 8\times 14$, $2\times 4\times 43$, and $2\times 13\times 16$.
Recently, Tardibuono et al.~\cite{TZDKJPUD2026} found a polyomino of area $106$ that folds into three boxes of sizes $1\times 1\times 26$, $1\times 2\times 17$, and $1\times 5\times 8$.
Xu et al.~\cite{XHSU2017} performed an exhaustive search and found nine polyominoes of area $30$, each of which folds into boxes of sizes $1\times 1\times 7$, $1\times 3\times 3$, and $\sqrt{5}\times \sqrt{5}\times \sqrt{5}$. The last box requires a fold along a diagonal of a $1\times 2$ rectangle to obtain an edge of length $\sqrt{5}$. We do not use this diagonal-folding technique; throughout this paper, we consider only foldings along the polyomino's grid lines.

The question of whether a polygon can fold into four or more distinct boxes has remained open.
In this paper, we answer this question affirmatively.
Our main theorems are as follows:
\begin{theorem}\label{th:four}
(1) There exists a polyomino of area $178$ that folds into four distinct boxes of size 
$1\times 1\times 44$, $1\times 2\times 29$, $1\times 5\times 14$, and $1\times 8\times 9$.
(2) There exist infinitely many polyominoes that can fold into four distinct boxes.
\end{theorem}
Our first four-box polyomino was discovered with the help of AI.
We analyze its structure and reduce its area from 532 to 178.
\begin{theorem}\label{th:five}
(1) There exists a polyomino of area $358$ that folds into five distinct boxes of size 
$1\times 2\times 59$, $1\times 3\times 44$, $1\times 4\times 35$, $1\times 8\times 19$, and $1\times 11\times 14$.
(2) There exist infinitely many polyominoes that can fold into five distinct boxes.
\end{theorem}
The first five-box polyomino we found was also discovered with the assistance of AI.
We analyze its redundant structure and reduce its area from 2128 to 358.
Both polyominoes in Theorems~\ref{th:four}(1) and~\ref{th:five}(1) can be generalized, yielding Theorems~\ref{th:four}(2) and~\ref{th:five}(2), respectively. 
Theorem~\ref{th:four}(2) implies Theorem~\ref{th:five}(2). 
Nevertheless, we include it because its construction is more flexible and may lead to polyominoes that fold into six or more boxes.
Materials and visualizations are available at \url{https://www.jaist.ac.jp/~uehara/boxes/}.

\section{Preliminaries}

As stated in the Introduction, a \emph{polyomino} is a connected union of unit squares in the plane, 
in which any two adjacent squares share a full edge.
The three-dimensional analogue is a \emph{polycube}. In this paper, however, we focus on rectangular cuboids with integer side lengths, which we call \emph{boxes}.
When we say a box of size $a\times b\times c$, we assume that each side length is a positive integer, that is, $a,b,c>0$.
Its six faces consist of two rectangles of size $a\times b$, two of size $b\times c$, and two of size $a\times c$.
We sometimes consider a special ``box'' with $c=0$.
In this case, we call it \emph{doubly covered rectangle} of size $a\times b$. 
That is, a doubly covered rectangle of size $a\times b$ consists of two coincident rectangular faces of size $a\times b$.
We say that a box $Q$ is \emph{covered} by a (connected) polyomino $P$ if $P$ can fold into $Q$ without gaps or overlaps. 
In this case, $P$ is called a \emph{net} of $Q$.
For a doubly covered rectangle, the two coincident faces are covered by the two layers of the folded sheet.

We assume that two boxes of sizes $a\times b\times c$ and $a'\times b'\times c'$ have the same surface area.
Equivalently, we have $ab+bc+ac=a'b'+b'c'+a'c'$. 
Without loss of generality, we assume that $a\le b\le c$ and $a'\le b'\le c'$.
Then the boxes are \emph{distinct} if and only if at least one of $a\neq a'$, $b\neq b'$, and $c\neq c'$ holds.

\section{Polyominoes that fold into four boxes}
\label{sec:four}

The first four-box net we found has area 532 and folds into $1\times 2\times 88$, $2\times 4\times 43$, $2\times 7\times 28$, and $2\times 13\times 16$.
It was found using the AI model GPT-6 Astra Ultra, starting from a briefing based on work on the problem with Claude Opus 5.5 (Anthropic).
Details are available on GitHub (\url{https://github.com/Sprite143/four-box-net-532}).
We now explain how to reduce its area from 532 to 178.

\subsection{Basic Ideas}


\begin{figure}[htbp]
  \centering
   \includegraphics[bb=0 0 2607 1034,width=0.8\linewidth]{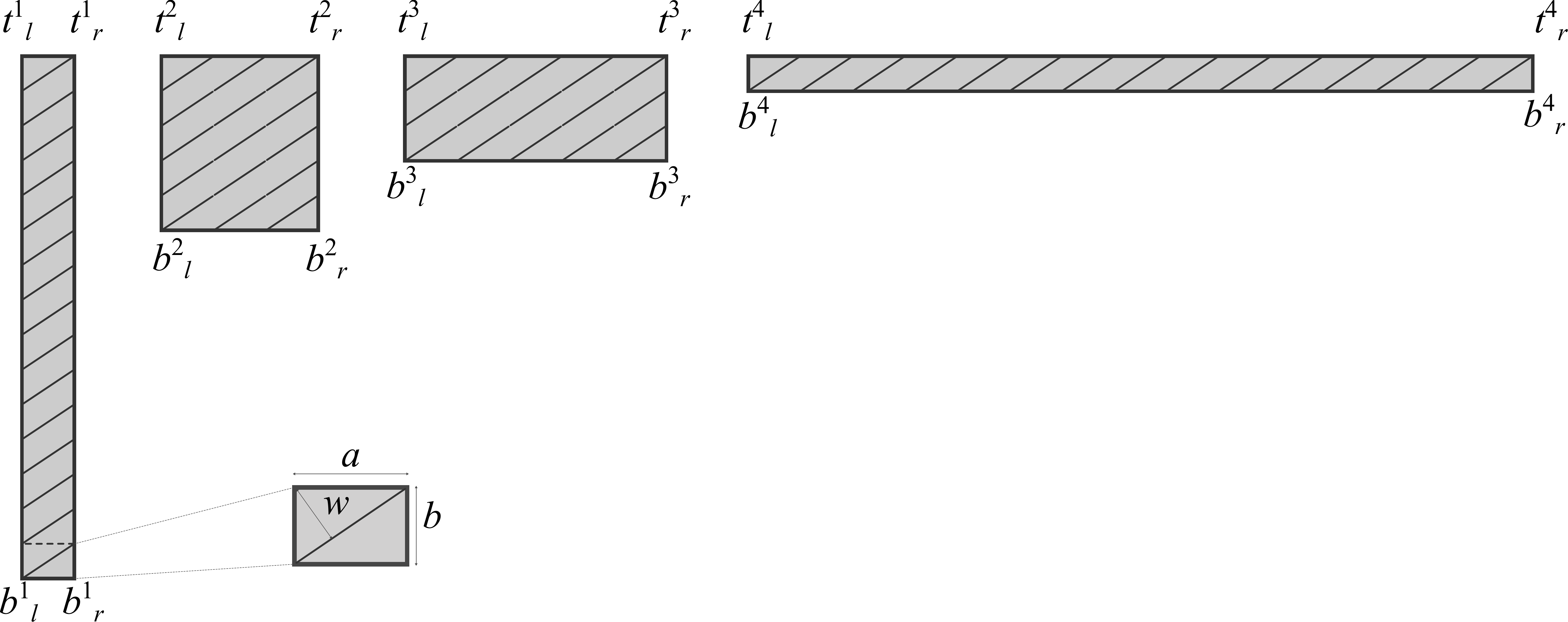}
   \caption{Doubly covered rectangles of size $a \times 15b$, $3a \times 5b$, $5a \times 3b$, and $15a \times b$ 
   folded from a common paper strip of size $\frac{ab}{\sqrt{a^2+b^2}}\times 30\sqrt{a^2+b^2}$.}
   \label{fig:double}
\end{figure}

\begin{figure}[htbp]
  \centering
   \includegraphics[bb=0 0 2734 1814,width=0.8\linewidth]{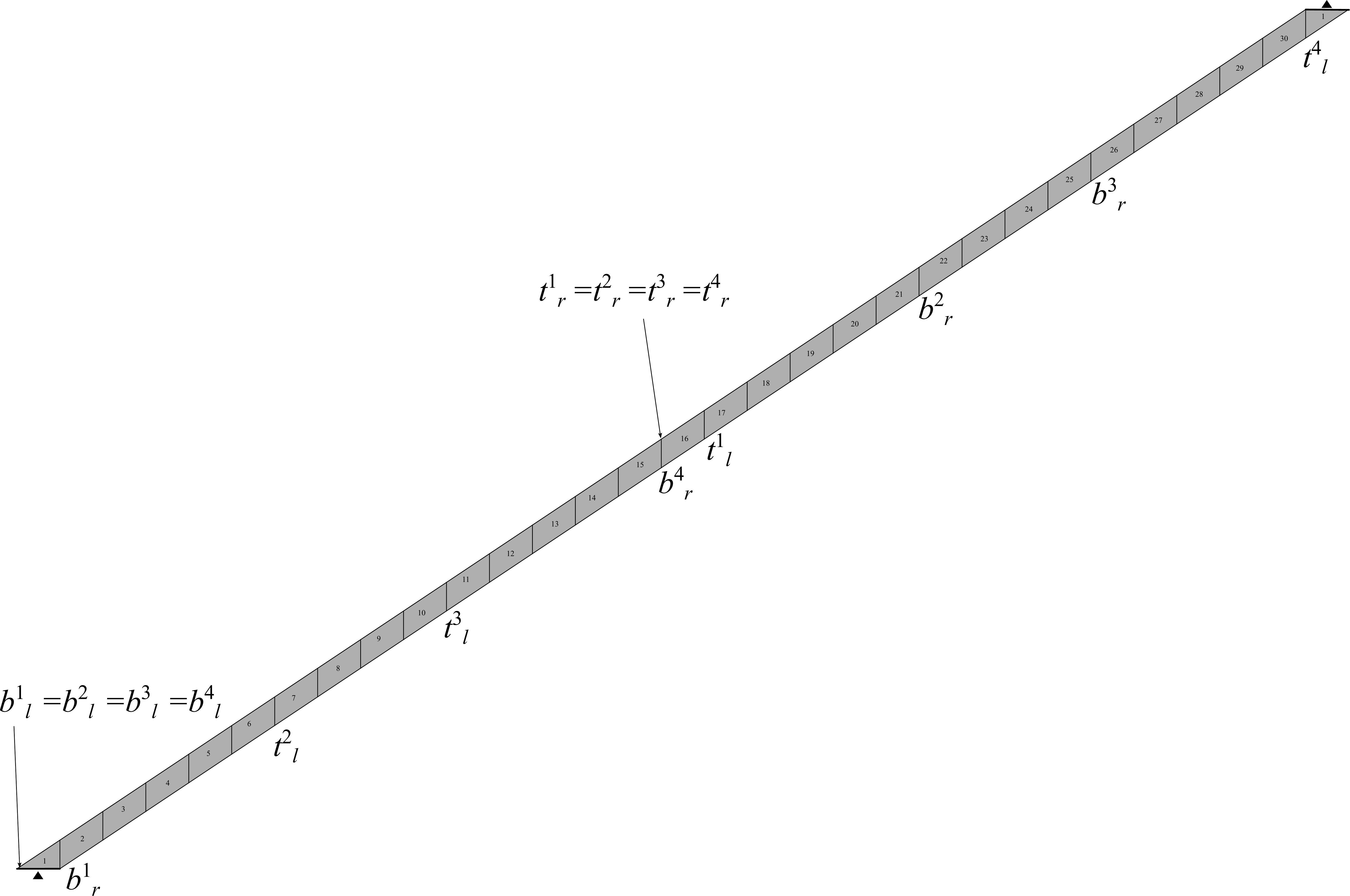}
   \caption{A common net for doubly covered rectangles of size $a \times 15b$, $3a \times 5b$, $5a \times 3b$, and $15a \times b$.}
   \label{fig:strip}
\end{figure}

We use two basic ideas. 
The first idea is that a certain family of rectangles can each be wrapped twice by the same paper strip, with the same width and length. Thus, a single rectangular strip can fold into multiple doubly covered rectangles. 
The same idea appears in \cite[Section 3.2]{AbelDemaineDemaineMatsuiRoteUehara2011}.
However, the authors note in their concluding remarks that its details are not explored.
We here give a technical lemma that is useful in constructing a common net for four distinct boxes.
\begin{lemma}
\label{lem:abpq-strip}
Let $p$ and $q$ be relatively prime odd integers at least 3, and let $a,b$ be 
real numbers greater than $1$.
Then there exists a paper strip of width $w=\frac{ab}{\sqrt{a^2+b^2}}$ and length $2pq\sqrt{a^2+b^2}$ that folds into 
four doubly covered rectangles of sizes $a\times bpq$, $ap\times bq$, $aq\times bp$, and $apq\times b$.
\end{lemma}
\figurename~\ref{fig:double} illustrates the case $p=3$, $q=5$, 
for arbitrary real numbers $a,b>1$.
Doubly covered rectangles of sizes $a\times 15b$, $3a\times 5b$, $5a\times 3b$, and $15a\times b$ can be folded from
the same paper strip of width $w=\frac{ab}{\sqrt{a^2+b^2}}$ and length $30\sqrt{a^2+b^2}$.
\begin{proof}
The centerline of the paper strip can be seen as a billiard trajectory in each rectangle.
Moreover, we can assume that these trajectories do not pass through any vertex of the rectangle.
In particular, every trajectory parallel to a periodic one is also periodic with 
the same period (or the length of the strip) \cite[Section~1.4]{MasurTabachnikov2002}.
We now verify that the width of the strip (equal to distance between two parallel trajectories) is always $w$
and the strip covers each rectangle exactly twice.

To compute $w$, let $L=\sqrt{a^2+b^2}$ and $\tau(x,y)=\frac{-bx+ay}{L}$,
so that $\tau$ is the signed coordinate perpendicular to the direction $(a,b)$.
We suppose that $(r,s)=(1,pq), (p,q), (q,p), \mbox{ or } (pq,1)$.
We consider a doubly covered rectangle as a flat torus obtained by identifying opposite sides of 
the unfolded rectangle of size $2ar \times 2bs$.
Then translations by the period vectors $(2ar,0)$ and $(0,2bs)$ on the flat torus change $\tau$ by
$-\frac{2abr}{L}\mbox{ and }\frac{2abs}{L}$, respectively. 
Since $\gcd(r,s)=1$, these changes generate $\frac{2ab}{L}\mathbb{Z}$.
Thus the transverse coordinate has period $2w$, where $w=\frac{ab}{L}$.

On the other hand, since each trajectory closes after length $2pqL$, 
the area of \(\mathcal{B}\) is $\operatorname{area}(\mathcal{B})=w\cdot 2pqL=2abpq$.
The rectangle has area $(ar)(bs)=abrs=abpq$, so the strip has twice the area of the rectangle. 

To establish the covering multiplicity, we count preimages directly.
Let $(X,Y)$ be an interior point that does not lie on the fold line.
Under the folding map from the unfolded flat torus to the rectangle, it has four preimages, 
represented by $(X,Y), (-X,Y), (X,-Y)$, and $(-X,-Y)$.
Their transverse coordinates, modulo $2w$, form two antipodal pairs: $u, -u, v, -v$.
An interval of length $w$ on a circle of circumference $2w$ contains exactly one point from each antipodal pair. 
Hence exactly two of the four preimages lie in $\mathcal{B}$. 
Every generic point of the rectangle is therefore covered exactly twice. 
The same construction extends to the boundary by continuity.

The argument above uses only $\gcd(r,s)=1$ and $rs=pq$. 
It therefore applies to $(r,s)=(1,pq), (p,q), (q,p), (pq,1)$.
In all four cases, the same strip, of width $\frac{ab}{\sqrt{a^2+b^2}}$
and length $2pq\sqrt{a^2+b^2}$, doubly covers the corresponding rectangle.
\end{proof}
\figurename~\ref{fig:strip} shows a common net for the four doubly covered rectangles in Figure~\ref{fig:double}.
When we align these rectangles to share the bottom-left vertex, by symmetry, the top-right vertex is also shared by them.
We note that the two ends of the strip are identified in each rectangle.
That is, in \figurename~\ref{fig:strip}, two edges indicated by two black triangles are glued in each of four ways of folding, 
 so the strip forms a loop of width $w$. 
This fact means that we can rotate vertices of some of four rectangles on the loop and 
shift the vertices on the boundary of the strip by cutting along the different positions.
We also note that each edge on the upper boundary of the strip is glued to another edge of opposite parity on the upper boundary,
and we have the same property on the lower boundary. 
That is, we can replace the unit straight line segment of length $\sqrt{a^2+b^2}$ on the boundary by 
any fixed unit curve unless it violates the connectivity.

\begin{figure}[htbp]
  \centering
   \includegraphics[bb=0 0 1212 771,width=0.8\linewidth]{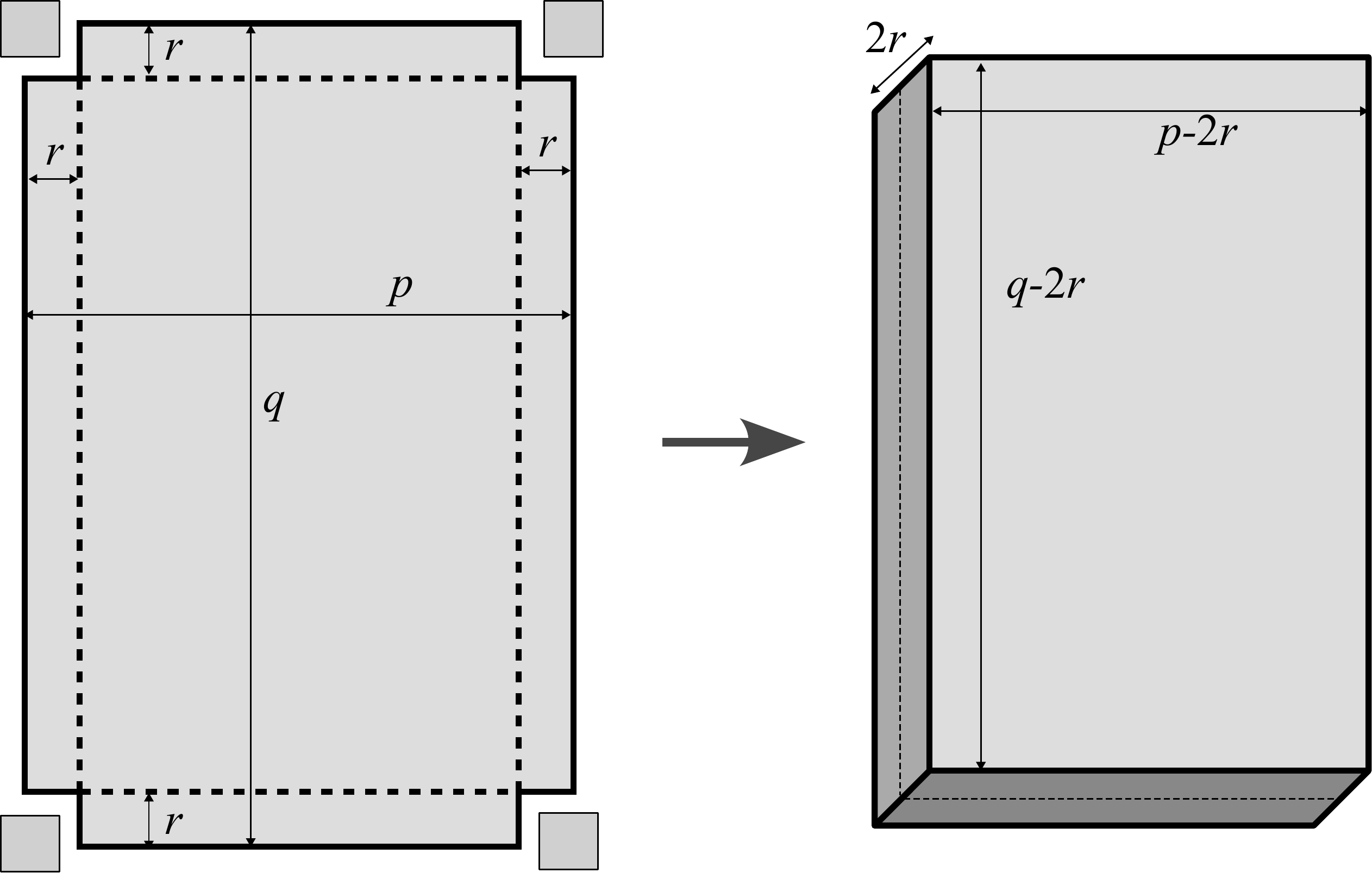}
   \caption{From a doubly covered rectangle to a box.}
   \label{fig:box}
\end{figure}

The second idea is to obtain a box from a doubly covered rectangle by cutting off its four corners and folding the remaining sheet (Figure~\ref{fig:box}). 
Specifically, given a doubly covered rectangle of size $p\times q$, cut off four squares of size $r\times r$ as shown in Figure~\ref{fig:box}. 
The remaining sheet folds into a box of size $(p-2r)\times(q-2r)\times 2r$ along the dotted lines.

\begin{figure}[htbp]
  \centering
   \includegraphics[bb=0 0 1281 1234,width=0.8\linewidth]{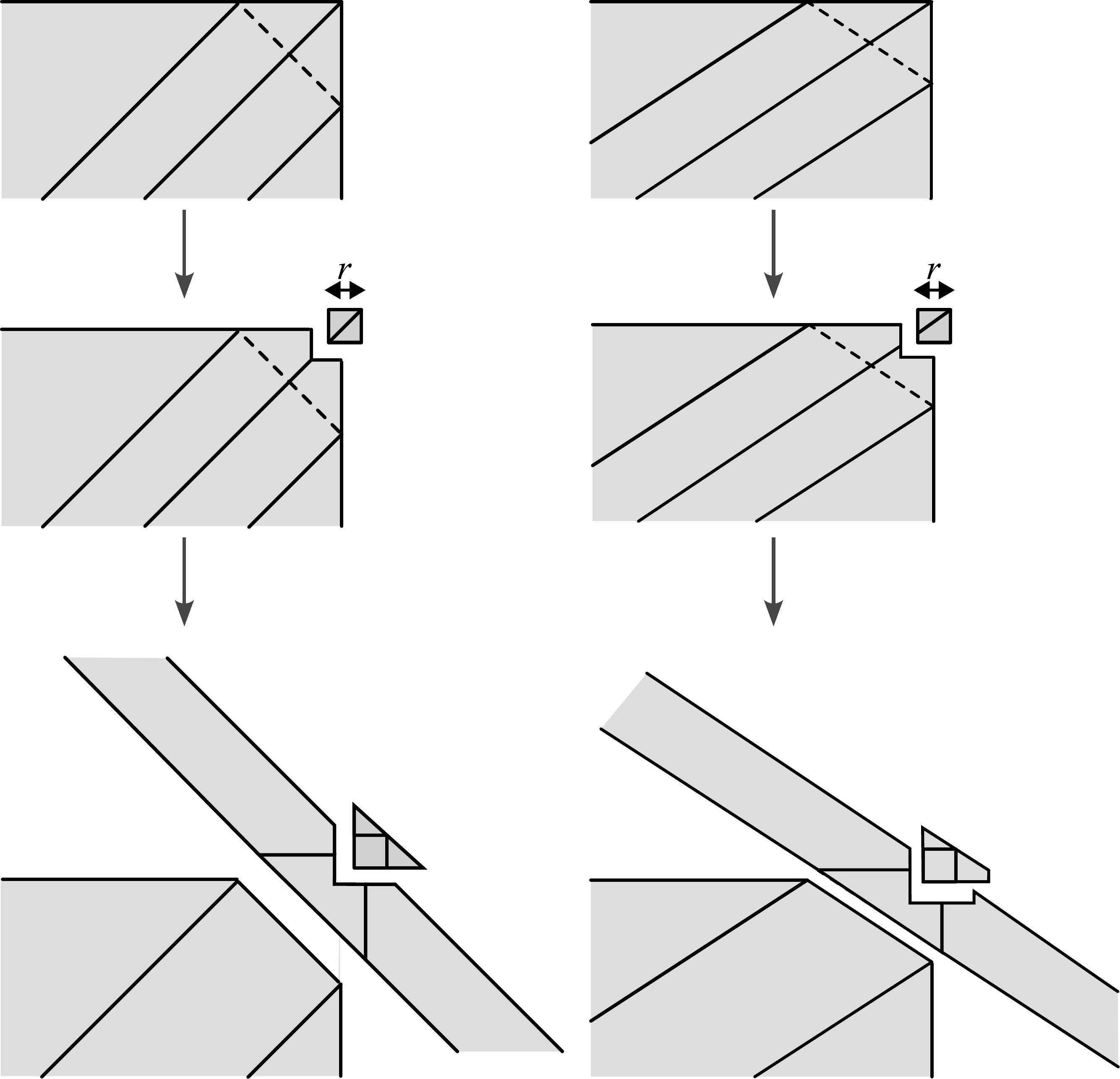}
   \caption{Shapes of the notch on the boundary of a strip.}
   \label{fig:notch}
\end{figure}

\begin{figure}[htbp]
  \centering
   \includegraphics[bb=0 0 2623 1245,width=\linewidth]{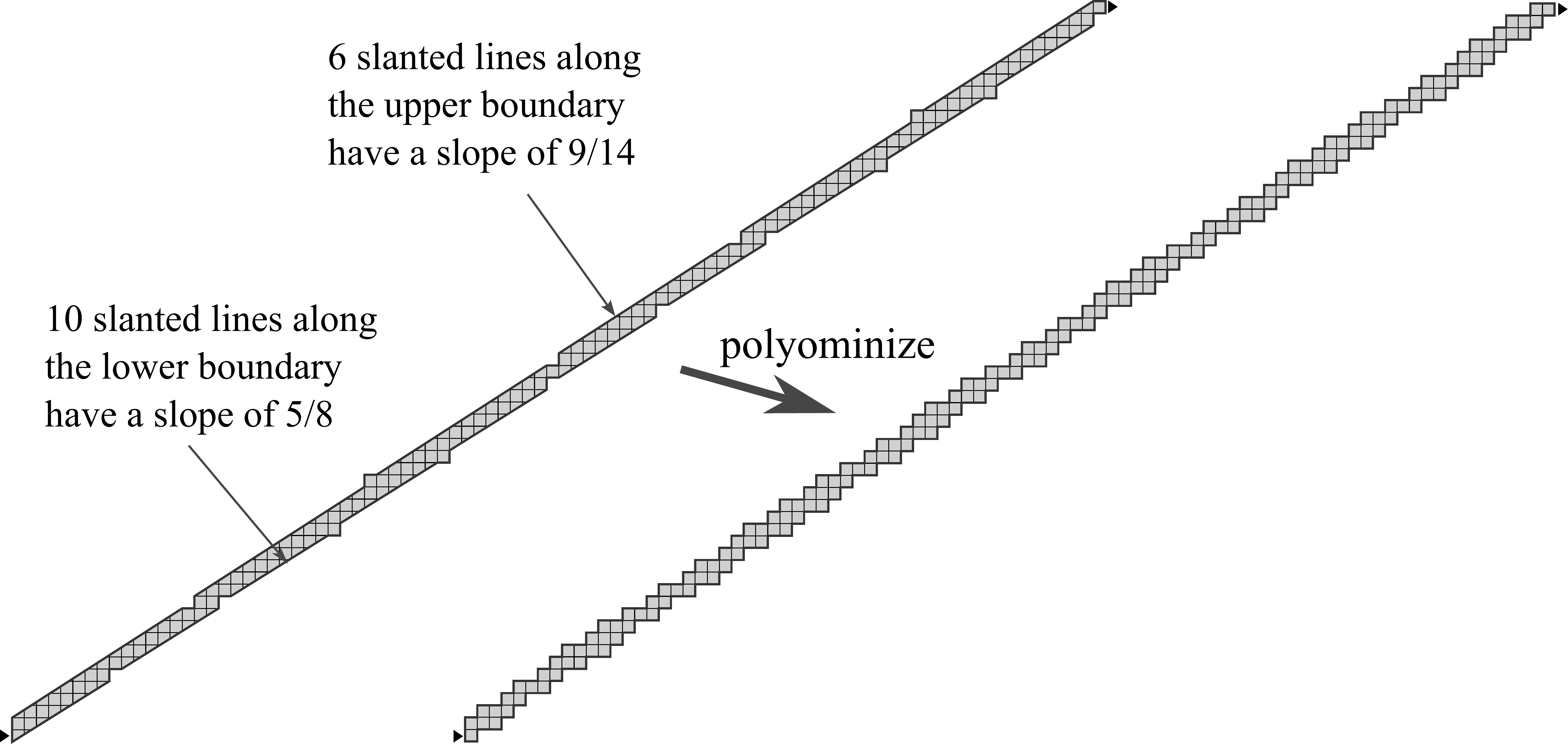}
   \caption{A polygon of area 178 that folds into four distinct boxes of size 
     $1\times 1\times 44$, $1\times 2\times 29$, $1\times 5\times 14$, and $1\times 8\times 9$.
     It can be transformed to a polyomino of area 178.}
   \label{fig:polygons-area178}
\end{figure}

\begin{figure}[htbp]
  \centering
   \includegraphics[bb=0 0 2629 1103,width=\linewidth]{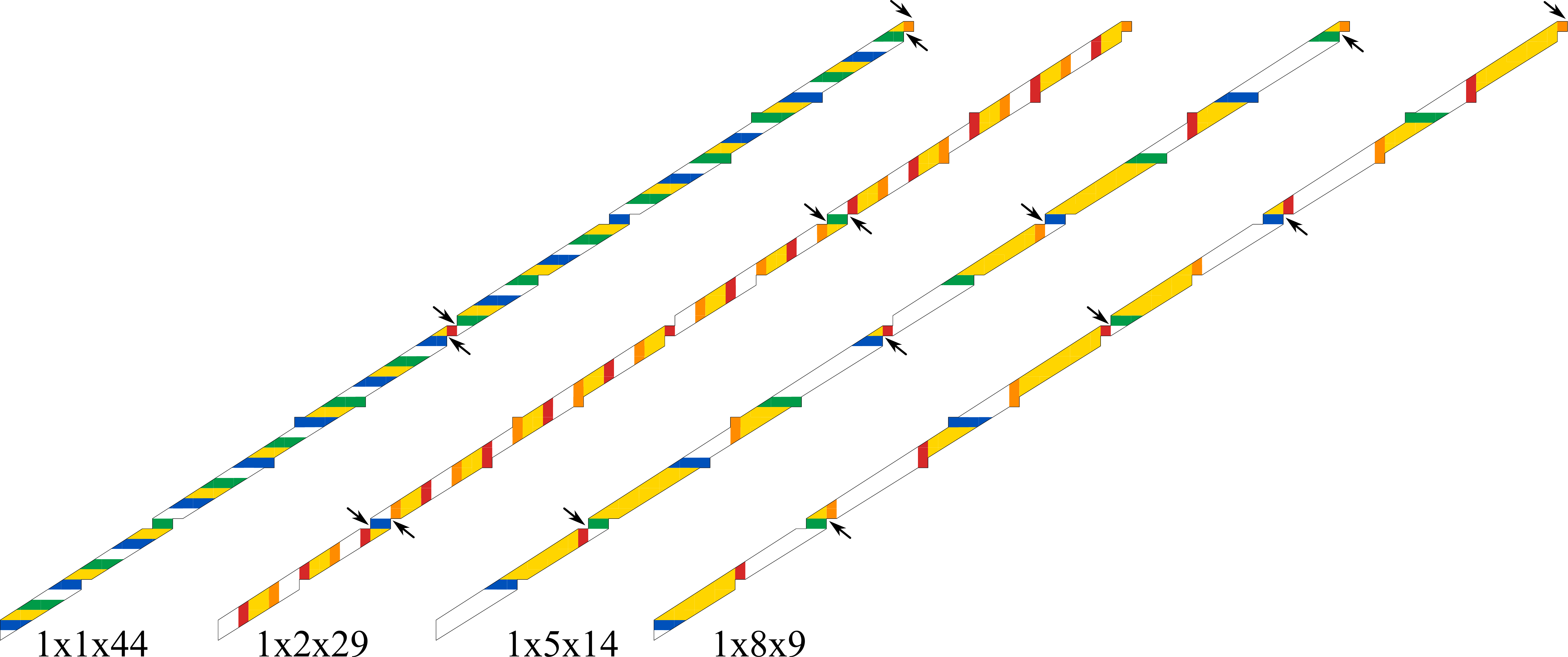}
   \caption{Four ways of folding of the polygon of area 178 in \figurename~\protect\ref{fig:polygons-area178} into four distinct boxes of size 
     $1\times 1\times 44$, $1\times 2\times 29$, $1\times 5\times 14$, and $1\times 8\times 9$.}
   \label{fig:polygons-area178-4ways}
\end{figure}

We now construct a paper strip that can fold into four distinct boxes. 
\begin{lemma}
\label{lem:abpq}
Let $p$ and $q$ be relatively prime odd integers at least 3, and let $a,b>1$ be real numbers.
Then there exists a polygonal net that folds into four boxes of sizes
$1\times (a-1)\times (bpq-1)$, $1\times (ap-1)\times (bq-1)$, $1\times (aq-1)\times (bp-1)$, and $1\times (apq-1)\times (b-1)$.
\end{lemma}
\begin{proof}
First, choose a pair of relatively prime odd integers $p$ and $q$ at least 3, and any real numbers $a,b>1$.
Let $\{R_1,R_2,R_3,R_4\}$ be the set of rectanbles of size $a\times bpq$, $ap\times bq$, $aq\times bp$, and $apq\times b$.
By Lemma \ref{lem:abpq-strip}, there is a paper strip (or loop) that wraps each rectangle twice, as shown in \figurename~\ref{fig:double}. 

Let $R_i$ be one of these rectangles. Each vertex $v$ of $R_i$ must lie on the boundary of the strip. 
We then cut off a square $s$ of size $1/2\times 1/2$ at $v$ from $R_i$ (namely, $r=1/2$ in \figurename~\ref{fig:box}).
Then $s$ itself is a doubly covered square.
As shown in \figurename~\ref{fig:notch}, the unfolding of $s$ is an isosceles right triangle when $a=b$, 
and a symmetric half of a square of size $1\times 1$ in general.

For each rectangle $R_i$, the notches occur in a fixed pattern along the strip, and 
the notch arrangements for $R_i$ and $R_j$ differ when $i\neq j$.
Therefore, when we fold the strip into $R_i$, 
the boundary of the strip must also have protrusions that fit the notches for the other boxes $R_j$.
Although we set $b^1_l=b^2_l=b^3_l=b^4_l$ in \figurename~\ref{fig:strip}, 
by shifting $b^2_l=b^4_l$ by $pq$, 
we can observe that all protrusions match with notches except ones folding to vertices of the corresponding boxes.
An example for $p=3$, $q=5$, $a=3$, and $b=2$ is given in \figurename~\ref{fig:polygons-area178-4ways}.
Each arrow marks a notch that folds to a vertex of the corresponding box.
In this polygon, every vertex of the polygon is placed on a grid line. 
To do that, we tilt each line segment between the vertices as shown in \figurename~\ref{fig:polygons-area178}.
As mentioned above, this tilting does not change the properties of this paper strip.
Extending this idea, replacing each of diagonal segments with suitably chosen zigzag lines 
transforms the strip into a polyomino as shown in \figurename~\ref{fig:polygons-area178}.
Since this paper strip has width $w=\frac{ab}{\sqrt{a^2+b^2}}$, it is not difficult to find the zigzag lines
without violating the connectivity of the strip of the same area.
\end{proof}

Now we prove Theorem \ref{th:four}:
\begin{proof}
The polygon of area $178$ and its polyomino version in Figure~\ref{fig:polygons-area178}
are obtained by setting $p=3$, $q=5$, $a=3$, and $b=2$ in Lemma \ref{lem:abpq}. 
This proves Theorem~\ref{th:four}(1).
Depending on the choice of $a,b,p,q$, the four boxes in Lemma~\ref{lem:abpq} need not be distinct. 
However, if, for example, $a,b,p,q$ are chosen to be positive integers satisfying $1<a<b<p<q$, 
then all four boxes are distinct.
This proves Theorem~\ref{th:four}(2).
\end{proof}

The strip shown in \figurename~\ref{fig:strip} forms a loop when its two ends are identified.
In each of the four foldings, the two edges marked by black triangles are glued together.
Each polygon and polyomino in this section inherits this property.
For example, in each of the four foldings of the polygon in \figurename~\ref{fig:polygons-area178}, 
the two edges marked by black triangles are glued together.
Therefore, cutting the loop at any position yields a common net for the four distinct boxes.
Consequently, for every choice of $a,b,p,q$ in Theorem~\ref{th:four}, 
we obtain infinitely many polygonal nets and $O(abpq)$ polyomino nets that fold into the same four distinct boxes.

\section{Polyominoes that fold into five boxes}

\begin{figure}[htbp]
  \centering
   \includegraphics[bb=0 0 985 774,width=\linewidth]{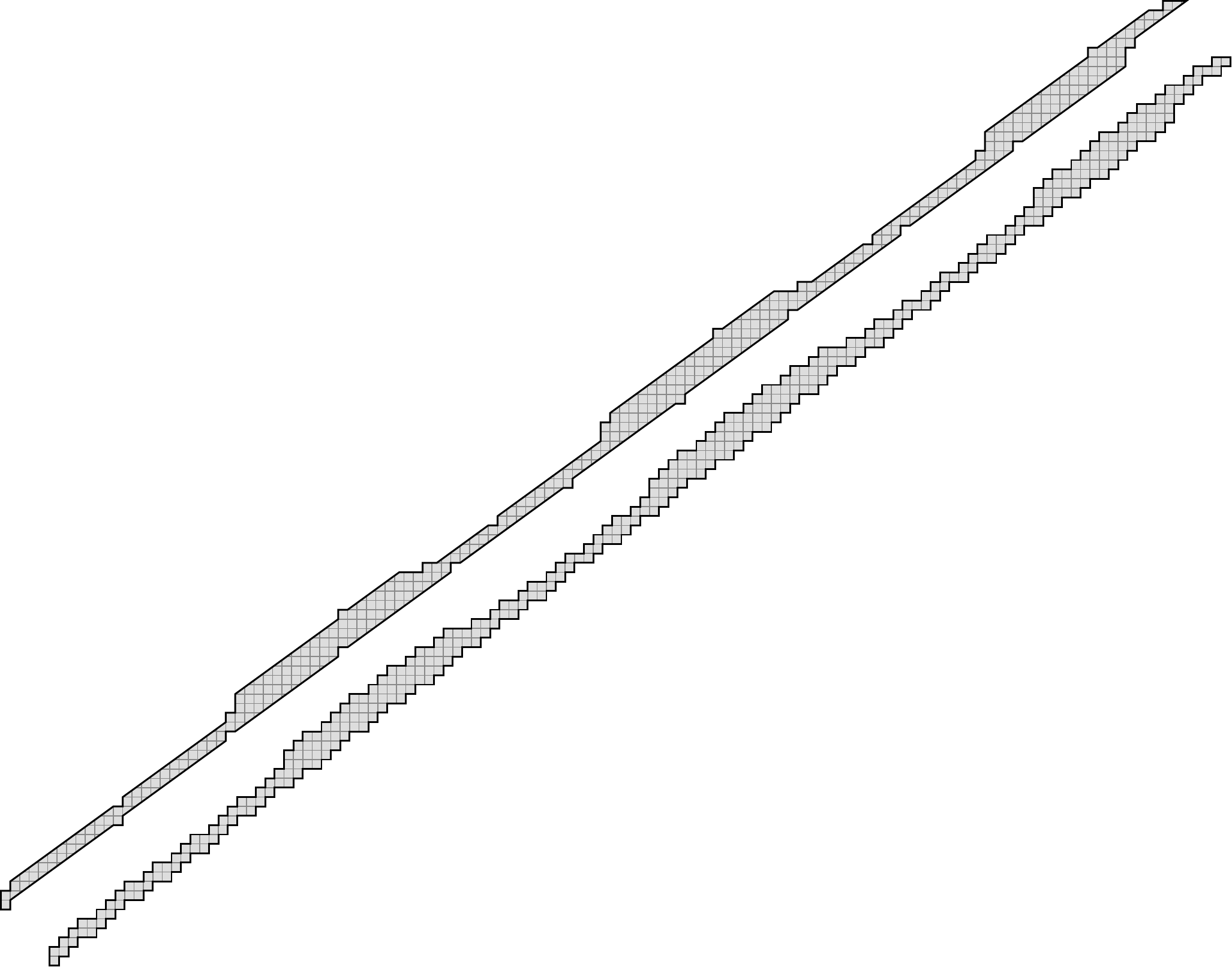}
   \caption{A polygon of area 358 which can fold into five distinct boxes of size 
     $1\times 2\times 59$, $1\times 3\times 44$, $1\times 4\times 35$, $1\times 8\times 19$, and $1\times 11\times 14$ and its polyomino version.}
   \label{fig:polygons-area358}
\end{figure}

The first five-box net we found has area 2128 and folds into 
five distinct boxes of size $2\times 4\times 176$, $4\times 8\times 86$, $4\times 14\times 56$, \underline{$4\times 16\times 50$}, and $4\times 26\times 32$.
It was found using GPT-6 Astra Ultra, starting from the four-box net with boxes of sizes 
$1\times 2\times 88$, $2\times 4\times 43$, $2\times 7\times 28$, and $2\times 13\times 16$ after subdividing each unit square into four squares of side length $1/2$.
Details are available on GitHub (\url{https://github.com/Sprite143/five-box-net-2128}).
Building on this result, we modified the area-178 polygon for four boxes in \figurename~\ref{fig:polygons-area178}
to obtain a polyomino of area 358 that folds into five boxes of sizes 
$1\times 2\times 59$, $1\times 3\times 44$, $1\times 4\times 35$, $1\times 8\times 19$, and $1\times 11\times 14$.
See \figurename~\ref{fig:polygons-area358} for the polygon and its polyomino version.
This polyomino proves Theorem~\ref{th:five}(1).

\begin{figure}[htbp]
  \centering
   \includegraphics[bb=0 0 693 702,width=\linewidth]{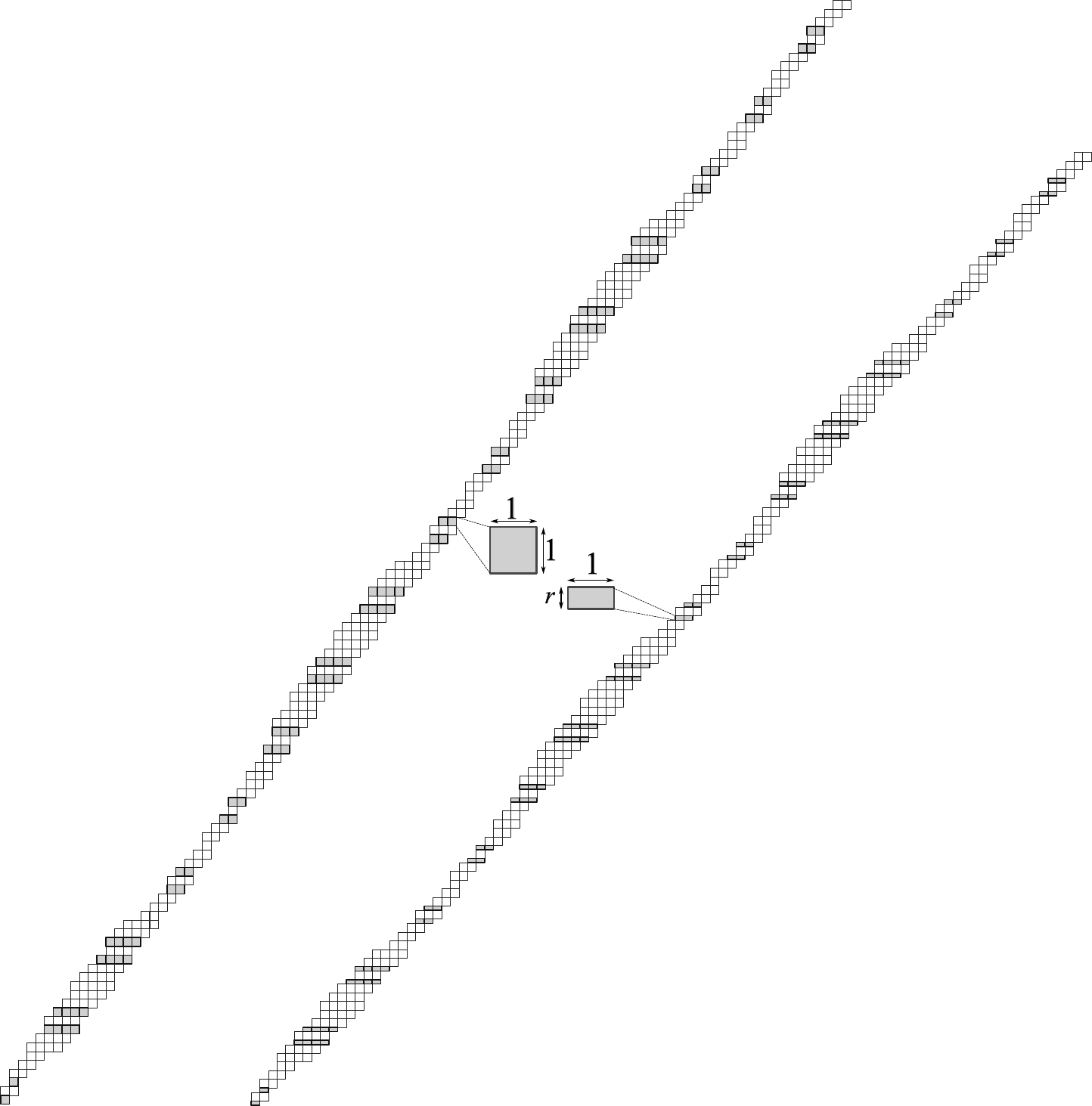}
   \caption{A polygon of area $268+90r$ that folds into five distinct boxes of sizes 
     $1\times 2\times (44+15r)$, $1\times (2+r)\times 44$, 
     $1\times 4\times (26+9r)$, $1\times 8\times (14+5r)$, and $1\times (8+3r)\times 14$ 
     obtained by replacing the gray unit squares with gray rectangles of size $1\times r$.}
   \label{fig:polygons-area358-S}
\end{figure}

We then used the AI model GPT-6 Luna to search for a set $S$ of unit squares in the polyomino that can be replaced by rectangles of size $1\times r$, for any real number $r>0$, while preserving the five-box folding property.
The model found a set $S$ of $90$ unit squares, shown in \figurename~\ref{fig:polygons-area358-S}.
Replacing these squares yields nets for five boxes with the following sizes: 
$1\times 2\times (59+15(r-1))=1\times 2\times (44+15r)$,
$1\times (3+(r-1))\times 44=1\times (2+r)\times 44$,
$1\times 4\times (35+9(r-1))=1\times 4\times (26+9r)$, 
$1\times 8\times (19+5(r-1))=1\times 8\times (14+5r)$, 
and $1\times (11+3(r-1))\times 14=1\times (8+3r)\times 14$.
For every positive integer $r$, the construction yields a polyomino; 
hence, it gives infinitely many polyominoes and proves Theorem~\ref{th:five}(2).\footnote{We here note that 
we have five distinct boxes when $r\neq 2$ and $r\neq 6$. When $r=2$ or $r=6$, two of five ways produce the same boxes.} 
\section{Concluding Remarks}

In this paper, we proved that infinitely many polyominoes can fold into five distinct boxes.
This extends previous results on common nets for two or three distinct boxes.
A major open problem is whether, for every positive integer $k$, there exists a common net (in particular, a polyomino) that folds into at least $k$ distinct boxes. In particular, can a common net fold into six distinct boxes?
For every positive integer $k$, there are at least $k$ distinct boxes with the same area~\cite[Theorem 3.1.1]{Uehara2020}.

Another open problem is to determine the minimum area for each value of $k$.
Mitani and Uehara~\cite{MitaniUehara2008} showed that the smallest polyomino known to fold into two distinct boxes has area $22$; an example is shown in Figure~\ref{fig:sample}.
The smallest known polyomino that folds into three distinct boxes has area $106$~\cite{TZDKJPUD2026}.
We give upper bounds of $178$ and $358$ on the minimum areas required for nets of four and five distinct boxes, respectively.
The area-consistency condition gives lower bounds of $46$ for three boxes ($1\times 1\times 11$, $1\times 2\times 7$, and $1\times 3\times 5$); 
$70$ for four boxes ($1\times 1\times 17$, $1\times 2\times 11$, $1\times 3\times 8$, and $1\times 5\times 5$); 
and $94$ for five boxes ($1\times 1\times 23$, $1\times 2\times 15$, $1\times 3\times 11$, $1\times 5\times 7$, and $3\times 4\times 5$), respectively.
Closing these gaps, as was done for $k=2$, remains open.

\bibliographystyle{plain}
\bibliography{references}

@incollection{MasurTabachnikov2002,
  author    = {Howard Masur and Serge Tabachnikov},
  title     = {Rational Billiards and Flat Structures},
  booktitle = {Handbook of Dynamical Systems},
  editor    = {Boris Hasselblatt and Anatole Katok},
  volume    = {1A},
  pages     = {1015--1089},
  publisher = {North-Holland},
  address   = {Amsterdam},
  year      = {2002},
  doi       = {10.1016/S1874-575X(02)80015-7}
}

@BOOK{Uehara2020,
	author="Ryuhei Uehara",
	title="{Introduction to Computational Origami}",
	publisher="Springer",
	year="2020",
}

@inproceedings{TZDKJPUD2026,
	Author = {Michael Tardibuono and Hanyu Zhang and Jenny Diomidova and Landon Kryger and Nicole Jacobus and Matthew Parker and Ryuhei Uehara and Erik D.~Demaine},
	Booktitle = {{The 38th Canadian Conference on Computational Geometry (CCCG 2026)}},
	Pages = {143--152},
	Title = {{Polyomino Nets Covering Three Different Boxes of Area 106 and Related Results}},
	Year = {2026}}

@book{DemaineORourke2007,
	Author = {E. D. Demaine and J. O'Rourke},
	Publisher = {Cambridge University Press},
	Title = {{Geometric Folding Algorithms: Linkages, Origami, Polyhedra}},
	Year = {2007}}

@unpublished{BiedlChanDemaineDemaineLubiwMunroShallit1999,
	Author = {T. Biedl and T. Chan and E. Demaine and M. Demaine and A. Lubiw and J. I. Munro and J. Shallit},
	Index = {CD},
	Month = {September 8},
	Title = {{Notes from the University of Waterloo Algorithmic Problem Session}},
	Year = {1999}}

@inproceedings{MitaniUehara2008,
	Author = {J. Mitani and R. Uehara},
	Booktitle = {{Canadian Conference on Computational Geometry (CCCG 2008)}},
	Pages = {39--42},
	Title = {{Polygons Folding to Plural Incongruent Orthogonal Boxes}},
	Year = {2008}}

@article{ShirakawaUehara2013,
	Author = {Toshihiro Shirakawa and Ryuhei Uehara},
	Journal = {International Journal of Computational Geometry and Applications},
	Number = {1},
	Pages = {65--71},
	Title = {{Common Developments of Three Incongruent Orthogonal Boxes}},
	Volume = {23},
	Year = {2013}}

@inproceedings{AbelDemaineDemaineMatsuiRoteUehara2011,
	Author = {Z. Abel and E. Demaine and M. Demaine and H. Matsui and G. Rote and R. Uehara},
	Booktitle = {{23rd Canadian Conference on Computational Geometry (CCCG 2011)}},
	Pages = {77--82},
	Title = {{Common Development of Several Different Orthogonal Boxes}},
	Year = {2011}}

@article{XHSU2017,
	author="Dawei Xu and Takashi Horiyama and Toshihiro Shirakawa and Ryuhei Uehara",
	title="{Common Developments of Three Incongruent Boxes of Area 30}",
	journal="COMPUTATIONAL GEOMETRY: Theory and Applications",
	year="2017",
	volume="64",
	pages="1--12",
	month="August",
	index=""}

\end{document}